




\let\miguu=\footnote
\def\footnote#1#2{{$\,$\parindent=9pt\baselineskip=13pt%
\miguu{#1}{#2\vskip -5truept}}}

     \def\=>{\Rightarrow}
\def \==> {\Longrightarrow}
\def\less{\backslash}



\def\ideq{\equiv}


\def\lto { {\raise1pt\hbox{$<$}} \!\!\!\! {\lower4pt\hbox{$\sim$}} }
\def\gto { {\raise1pt\hbox{$>$}} \!\!\!\! {\lower4pt\hbox{$\sim$}} }



\def\+{{\amalg}}

\def\singlespace{\baselineskip=12pt}
\def\sesquispace{\baselineskip=18pt}

\raggedbottom

\magnification=\magstep1

\hsize=6 true in
 \hoffset=0.27 true in
\vsize=8.5 true in
 \voffset=0.28 true in

\sesquispace

\rightline {SU-GP-93-12-1}

\bigskip

\centerline{\bf QUANTUM MECHANICS AS QUANTUM MEASURE THEORY}


\bigskip
\singlespace
\centerline {\it Rafael D. Sorkin}
\smallskip
\centerline {\it Department of Physics}
\centerline {\it Syracuse University}
\centerline {\it Syracuse, NY 13244-1130}
\smallskip
\centerline {\it  internet address: rdsorkin@mailbox.syr.edu}

\bigskip
\singlespace
\leftskip=1.5truecm\rightskip=1.5truecm     
\centerline{\bf Abstract}
\medskip
\noindent

The additivity of classical probabilities is only the first in a
hierarchy of possible sum-rules, each of which implies its successor.
The first and most restrictive sum-rule of the hierarchy yields
measure-theory in the Kolmogorov sense, which physically is appropriate
for the description of stochastic processes such as Brownian motion.
The next weaker sum-rule defines a  {\it generalized measure
theory} which includes quantum mechanics as a special case.  The fact
that quantum probabilities can be expressed ``as the squares of quantum
amplitudes'' is thus derived in a natural manner, and a series of
natural generalizations of the quantum formalism is delineated.
Conversely, the mathematical sense in which classical physics is a
special case of quantum physics is clarified.  The present paper
presents these relationships in the context of a ``realistic''
interpretation of quantum mechanics.

\bigskip
\leftskip=0truecm\rightskip=0truecm         


\sesquispace
\bigskip\medskip

 An attitude toward Quantum Mechanics which is suitable for quantum
gravity in general, and for its application to cosmology in particular,
is not so easy to find.  Understanding the early universe requires us to
reason about a time in the distant past in which observers in the ordinary
sense of the word can hardly have been present.  For such a situation, a
philosophically   ``realistic''
attitude toward quantum mechanics would seem to be more effective than
one based on operators which must find their physical meaning in terms of
``measurements''.  If the reality in question is taken to be
something with a ``spacetime'' character (such as a Lorentzian 4-geometry, or
some more fundamental discrete structure like a causal set)\footnote{*}
{%
Quantum gravity seems to demand a ``spacetime approach'' for more than one
reason, including the need to incorporate topology-change [1],
and the evident impossibility of making sense of continuous Hamiltonian
evolution for a discrete structure such as a causal set (for the latter
see e.g. [2]).
},
then the simplest description of its dynamics will be directly in terms of
probabilities of spacetime alternatives, rather than indirectly in terms of
operators and Hilbert spaces.  For this reason, the mathematics of the
``sum-over-histories'' is more akin to measure theory than to (say) lattice
theory or the theory of $W^*$-algebras.  Quantum dynamics in such a
formulation appears as a kind of generalization of the theory of stochastic
processes, rather than (directly) of classical mechanics.

To an untutored mind, however, the formal rules of the path-integral
scheme, could seem unnatural and contrived.  Why are probabilities {\it
squares} of amplitudes; why are they expressed most naturally in terms of
{\it pairs} of paths rather than individual paths?  (cf. [3]
[4] [5] [6])  We will see that a possible answer to
this question emerges if one places quantum mechanics in a still more
general context by asking whether quantum probabilities preserve any of
the additivity of classical ones.  Let us begin by considering, not the
ubiquitous two-slit diffraction experiment, but a generalization which I
will call the three-slit experiment.

\bigskip
\noindent{\bf The three-slit experiment }
\nobreak

  Imagine an experiment in which an electron (say) passes through any
one of three slits and impinges on an array of detectors.  Imagine that
you record the diffraction pattern with all three slits open, and then
repeat the procedure with some of the slits blocked off.  In all, you
can obtain in this way a total of eight diffraction patterns.  Now
superimpose the eight patterns, using a plus sign when an odd number (3
or 1) of the slits were open, and a minus sign when an even number (2 or
0) were open.  What will be the result?  Remarkably, you will always get
zero, as can be straightforwardly demonstrated.  Were the electron a
classical particle, you would also get zero, since each of the three
slits would contribute twice with a positive sign and twice with a
negative one.  In this sense, quantum randomness preserves something of
the classical additivity of probabilities.

One can go further and imagine diffraction experiments with four or more
slits.  For each case beyond two slits the analogous superposition will
again yield zero, but it turns out that these subsequent relations yield
nothing new, each of them being logically contained in the three slit
relation.  I will describe this hierarchy of sum-rules more carefully
below, but first I want to sketch the interpretive framework in which I
would propose to situate them.

\bigskip
\noindent{\bf  ``Quantum materialism'' and the quantum measure}
\nobreak

In accord with the above introductory remarks, we do not want to base
the interpretation of the generalized probabilities we will be dealing
with on some undefined concept of ``measurements made by human
observers''.  Instead I would propose a framework in which the ontology
or ``kinematics'' and the dynamics or ``laws of motion'' are as sharply
separated\footnote{*}
{
a separation which, presumably, must be overcome by the further
development of physical theory.
}
from each other as they are in classical physics (see [5] [4] [7]).  In
fact I will take the attitude that the ontology of quantum mechanics is
identical to that of classical realism (in a spacetime mode), according to
which the world {\it is} a single ``history''.  How this ``history'' is
actually structured is for science to find out, and the chosen kinematics
has varied from theory to theory.  According to the choice one has made,
the world might be described as a collection of world lines, a
spacetime-geometry (= diffeomorphism equivalence class of Lorentzian
metrics), a causal set, or something else.  But in any case, {\it all
meaningful statements of fact can by assumption be reduced to assertions
about this one existing history}.  Notions such as state-vectors and
observables never appear, except for the sake of computational
convenience.\footnote{*}
{
The account just given leaves open the question whether the history
should be thought of as existing ``timelessly'' like a painting, with
the entire future already laid out, or as a developing, incomplete thing
which ``grows at its tips like a tree''.  I personally believe the
latter, but nothing in the discussion which follows will require that a
choice be made.
}

Where quantum theory differs from classical mechanics (in this view) is in
its dynamics, which of course is stochastic rather than deterministic.  As
such, the theory functions by furnishing probabilities for sets of
histories.  More formally, it associates to a set $A$ of histories a
non-negative real number $|A|$, which I will call its {\it quantum measure}
$|A|$; and it is this measure that enters into the sum-rules we will be
concerned with.

In the {\it two}-slit experiment, for example, the probability that a
particular detector will register the arrival of the electron is
(proportional to) the measure $|C|$ of the set $C$ of all electron world
lines which in fact pass close enough to that detector to trigger it.
When we contemplate also blocking off one or the other slit, there are
(for a fixed detector) three sets of histories to consider: the set $A$ of
histories which arrive at the detector after traversing the ``first''
slit, the corresponding set $B$ for the ``second'' slit, and the
original set $C = A \amalg B$, the disjoint\footnote{$^\dagger$}
{
ignoring the possibility that the electron's trajectory winds around in
such a manner as to go through both slits.
}
union of $A$ and $B$.
It is of course characteristic of quantum probability  that the
{\it interference term}
$$
  I(A,B) := |A \amalg B| - |A| - |B|
$$
between the slits is not zero.  The surprising thing (once one has gotten
used to the fact of interference itself) is that this violation of the
classical probability sum-rules is in a certain sense so mild, since the
corresponding sum-rule for three alternatives remains valid.

In any case, the important thing from the standpoint of interpretation
is that the electron follows one and only one path, not somehow two at
once.  If probabilities are involved, it is only because the path is not
determined in advance, just as it is initially undetermined in a
classical stochastic process.

Given the failure of the sum rule $I(A,B)=0$, it is clear that quantum
probabilities cannot be interpreted in the same manner that classical
ones are wont to be interpreted, in terms of (actual or fictitious)
ensemble frequencies.  How they should be interpreted is a question to
which I will return briefly below, and more at length in
another place [8].  Here, my main purpose is to discuss the sum-rules
themselves.

\bigskip
\noindent{\bf Quantum measure theory and its generalizations }
\nobreak

What ordinarily makes it difficult to regard quantum mechanics as in
essence a modified form of probability theory, is the peculiar fact that it
works with complex ``amplitudes'' rather than directly with probabilities,
the former being more like square roots of the latter.  To put this
peculiarity in context, consider the following series of symmetric
set-functions, which generalize the interference term $I(A,B)$ introduced
above.  (Notice that all the sets $A$, $B$, $C\cdots$ which occur here are
mutually disjoint.)
$$\eqalignno{
  I_1(A)     &\ideq |A|                                       & \cr
  I_2(A,B)   &\ideq |A \amalg B| -|A| -|B|                    & \cr
  I_3(A,B,C )&\ideq |A \amalg B \amalg C| -
                  |A \amalg B| - |B \amalg C| -|A \amalg C|
                     + |A| +|B| + |C|,                         & \cr
}$$
or in general,
$$\eqalignno{
    I_n(A_1,A_2,\cdots,A_n) &\ideq  |A_1 \amalg A_2 \amalg \cdots A_n|&\cr
            & - \sum | (n-1) sets | + \sum | (n-2) sets | \cdots&\cr
            & \pm \sum\limits_{j=1}^{n} |A_j|           & (1)
}$$
These expressions are related sequentially in a simple manner expressed
by the following lemma, whose straightforward inductive proof will be
given elsewhere.

\noindent
{\bf Lemma}\ \  $I_{n+1}(A_0, A_1, A_2, \cdots, A_n) =
                            I_{n}(A_0 \amalg A_1, A_2, \cdots, A_n)
                -  I_{n}(A_0, A_2, \cdots, A_n)
                -  I_{n}(A_1, A_2, \cdots, A_n)$

  For each $n$ one obtains a possible sum-rule by setting $I_n$ to
zero.  It is an immediate consequence of the lemma that the $n^{th}$
such sum-rule entails the $(n+1)^{st}$.  Hence the sum-rules form a
hierarchy of ever decreasing strength.  The first sum-rule in the
hierarchy, $I_1 = 0$, trivializes the measure and is therefore
uninteresting.  The second expresses precisely the additivity of
classical measure theory, or equivalently the additivity of classical
probabilities, when they are regarded as set-measures in the Kolmogorov
manner.  Accordingly, the third sum-rule, $I_3(A,B,C) \ideq 0$, defines
a generalization of measure theory which preserves most, but not all, of
the additivity of classical probabilities.  This is the level of quantum
measure theory.  The fourth and higher sum-rules define still more
general forms of measure theory, which may be regarded as natural
extensions of quantum mechanics.

A second immediate consequence of the lemma is the fact that $I_{n+1}$
vanishes if and only if $I_n$ is ``additive'' in each argument, given
the mutual disjointness of all its arguments.  Thus each sum-rule is
associated with a kind of
multilinearity (really multi-additivity) of the
function which measures the failure of the next stronger sum-rule to
hold.  At the quantum level, specifically, we learn that $I_2$ is
bi-additive, and we will see that the peculiar quadratic
relationship between quantum amplitudes and probabilities corresponds directly
to this feature of $I_2$.  In the next generalization beyond the quantum
level, tri-additivity would take the place of bi-additivity and (insofar as
something like a quantum state-space were relevant at all) some sort of
trilinear form associated to $I_3$
 would presumably replace the familiar inner
product of quantum Hilbert space.  In what follows, however we will limit
ourselves to the quantum case as defined by the $n=3$ sum-rule,
$$
    |A \amalg B \amalg C| -  |A \amalg B| - |B \amalg C| -|A \amalg C|
     + |A| +|B| + |C| = 0.           \eqno(2)
$$

Given this sum-rule, we can, as just pointed out, conclude immediately from
the lemma that   $I_2$ is bi-additive in the sense that
$$
    I(A \amalg B, C) = I(A,C) + I(B,C),  \eqno (3)
$$
whenever $A$, $B$ and $C$ are mutually disjoint.  (Henceforth, I will
usually omit the subscript `2' from `$I_2$'.)  Full bi-additivity of $I$,
however
would require this same equality even when $C$ overlaps $A$ or $B$,
a situation in which $I$ has not even been defined.  This raises the
obvious question whether we can extend the definition of $I(A,B)$
to general arguments in such a way as to preserve its bi-additivity.

Supposing such an extension to have been made, consider the combination
$I(A\+B,A\+B)$.  Expanding it out via bi-additivity and
rearranging, we find that, for disjoint subsets  $A$ and $B$,
$$
  2\, I(A,B) = I(A\+B,A\+B) - I(A,A) - I(B,B),
$$
which, on comparison with the defining equation for $I_2$,
strongly suggests the identification
$$
     I(X,X) = 2 \, |X|.                    \eqno (4)
$$ If we adopt this as the value of $I$ on equal arguments, then its
value for arbitrary arguments is completely determined by bi-additivity;
and a short computation which will appear elsewhere confirms that the
resulting definition of $I(A,B)$ is self consistent.  The end result is
that $I$ can
be expressed in terms of the quantum measure $|\,\cdot\,|$ in several
equivalent forms, of which two are the following.
$$
    I(A,B) = |A \cup B| + |A\cap B| - |A\less B| - |B\less A|, \eqno (5)
$$
$$
    I(A,B) = |A \Delta B| +|A| +|B| -  2\, |A\less B|  - 2\, |B\less A|.
$$
(In these equations the symbol `$\less$' denotes set-theoretic difference
and `$\Delta$' denotes symmetric difference.)

We thus conclude that any generalized measure obeying the quantum sum rule
(2) can be expressed in the form $|X| = I(X,X)/2$, where $I$ is the
bi-additive, real-valued set function of (5).  Conversely, we could begin
with such a set-function whose diagonal values are all non-negative, and
use it to define a quantum measure $|\cdot|$ obeying the sum rule (2).  This
is what is normally done, with $I$ taken to be what reference [9] would
call (twice the real part of) the ``decoherence functional''.  The
postulate that quantum probabilities should be derived from such
a bi-additive function can thus be replaced by the assumption that
they obey the fundamental sum-rule (2).

For completeness, let me conclude this section by sketching the way that
the ordinary non-relativistic quantum mechanics of point particles fits
into this framework [4].  Also, since none of our discussion has attempted
to address the measure-theoretic technicalities associated with continuous
spaces of histories, let me pretend that the set of all possible particle
paths has finite cardinality.  Then the measure of any set
$A = \{x,y,\cdots,z\}$ of paths can be formally expressed as
$|A| = ``(1/2) I(x+y+\cdots z,x+y+\cdots z)$'',
which is to be evaluated by expanding out the sums via bilinearity and
interpreting $I(x,y)$ as $I(\{x\},\{y\})$.  To complete the construction we
must take $I(x,y)$ to be essentially $e^{-i S(x)} e^{i S(y)} + (complex\
conjugate)$, where $S(x)$ is the Action of the path $x$.

Actually the true expression is somewhat more complicated than
this, and requires the introduction of a ``truncation time'' $T$ lying
to the future of the span of time to which the properties defining $A$
refer [4].  The actual rule then
involves paths truncated  to  time $T$, and the expression
for $I(x,y)$ acquires a delta-function which ``ties together'' the final
endpoints of the truncated paths.  With the convention that $x$ and $y$
now represent such truncated paths, $|A|$ is given finally as the sum over
all $x$ and $y$  belonging to $A$ of the expression,
$$
  I(x,y) = \delta(x(T),y(T)) \; e^{-i S(x)} e^{i S(y)} + (complex\ conjugate).
$$
The import of this rule can also be rendered by the statement that the
measure $|A|$ of a set of trajectories is the norm-squared of the wave
function which is produced at time $T$ by restricting the path-sum to
the set of paths  belonging to $A$.  This last statement is
recognizably the standard quantum probability rule, as expressed in
sum-over-histories language.

\bigskip
\noindent{\bf  Final remarks on interpretation and some open questions}
\nobreak

Although we have succeeded in tracing the main traits of the quantum
formalism to the fundamental sum-rule (2) for the quantum measure, we
have only raised, without settling, the question of how this measure or
``quantum probability'' is to be interpreted physically.  That question
entrains far too many issues for a short manuscript to deal with, but
the present paper would be incomplete without at least some indication
of an answer.

With a frequency interpretation of the measure unavailable, it is natural
to adopt the attitude that locates the predictive content of a (classical
or quantum) probabilistic theory in the assertion that events of
sufficiently small measure, for all practical purposes, do not occur at
all: they are {\it precluded} events in the language of [10].  The meaning
of the measure, then, would be that the true history will not (or ``almost
never'') belong to a precluded set $A$.  The trouble with this rule is
that, if used without the requisite tact, it leads to mutually conflicting
predictions.  In Feynman's well-known exposition of the two-slit
experiment, for example, the partition of the histories which properly
comes into play depends on what question we are asking, and not all
questions can be simultaneously valid.  Here, asking a question is not
something ``mental'', but something ``material'' like putting a detector in
place, and paradox is avoided because the questions leading to conflicting
preclusions are not all realizable within a single experiment.

However satisfactory such a resolution might be for most practical
purposes, it threatens to bring back the same subjectivity and
human-centeredness which all along we have been at pains to reject.  We can
retain objectivity, I believe, by abstracting from the idea of measurement
the idea of correlation, and limiting the application of preclusion to
situations where an appropriate type of correlation occurs.  Briefly, the
idea is that, when variables pertaining to spacelike-separated regions
become correlated (in the sense that the correlation-breaking possibilities
correspond to sets of histories of small or zero measure), then the failure
of the correlation is precluded.  (For example, if one of a pair of
electrons with anti-correlated spins traverses the ``$\sigma_z = +1$'' beam
of a Stern-Gerlach analyzer, then the other must traverse the
``$\sigma_z=-1$'' beam of its analyzer, assuming these events are spacelike
separated.)  Moreover, we predict in such a situation that, if one of the
{\it correlated} possibilities $P$ is itself of negligible measure,
then it also
is precluded (i.e. we predict that the true history almost certainly does
not belong to $P$).  Notice that these predictions are statements about the
history itself, not just about ``what we would find if we observed the
variables in question''.

The double predictive principle just enunciated seems to suffice for using
quantum mechanics in the way we use it, without leading to any obvious
contradictions.  Unfortunately it does lead to some unobvious
contradictions, but all those that I know of can be excluded by a small
refinement of the predictive scheme.  The refinement in question, and
further details of the resulting prescription will be discussed in another
place [8].  Here there is space only to raise a few further questions
which are naturally suggested by the preceding development.

The first question is whether some further axiom of general validity can or
should be added to our basic sum-rule (2).  It is a feature of standard
quantum mechanics (and also of classical probability theory) that the
measure of a set of histories $A$ is unaltered when a disjoint set of
measure zero is adjoined to it; in particular the union of two disjoint
sets of measure zero will also have zero measure.  Since this property is
natural, and turns out to be important for the analysis of reference [8],
it appears reasonable to adopt it as a further condition on the measure
$|\,\cdot\,|$.  (For a related proposal in the language of ``decoherence''
see [11].)

A second question is whether there is a sense in which a process
governed by quantum measure theory can be called Markovian.  With the
idea of amplitude not taken for granted, it is not obvious that the
answer is yes, but if it is, then it would be interesting to find
necessary and sufficient conditions  in
terms of  $I(\cdot,\cdot)$
for a quantum measure to be Markovian.
  Such conditions ought to clarify what makes
ordinary quantum mechanics special among possible solutions of (2), and
might suggest novel generalizations of Hamiltonian evolution as well.

A third question is whether the introduction of the ``truncation time''
$T$ in the previous section was really needed.  In the classical theory
of stochastic processes, such a truncation of the histories is
unnecessary, and the similarity of that theory to the present
formulation offers hope of avoiding it in ``quantum stochastic
mechanics'' as well.  Success in this could be crucial for quantum
gravity, which lacks any background time with respect to which
truncation could be carried out (see however [4]).

The asking of these last two questions highlights the fact that not all of
the somewhat elaborate details of the construction of $I(A,B)$ in the
previous section are included in the simple sum-rule (2).  Some of them are
related instead to the Markovian character of non-relativistic quantum
mechanics, and beyond that, to the unitary evolution it embodies.  However,
non-unitary and non-Markovian evolution is the rule in open systems
(e.g. systems coupled to ``reservoirs''), so that the greater generality of
what we have here called ``quantum measure theory'' already has important
application.  Moreover, it appears unlikely that unitarity will have
fundamental meaning for quantum gravity, and one may suspect that something
at least as general as full quantum measure theory will be needed for that
theory, if not some still more general dynamical framework, perhaps
corresponding to one of the higher sum rules described in this paper.

Finally, let us return for a moment to the three-slit experiment with
which we began.  If some more general form of dynamics than quantum
mechanics is at work in nature, it should show itself in a failure of
the sum rule (2) for which the three-slit discussion is a prototype.
Thus, any situation in which three distinct alternatives can
interfere offers a potential ``null test'' of the validity of quantum
mechanics.  It might be worthwhile looking for experimentally realizable
situations of this type where, unlike with ordinary diffraction, the
satisfaction of the test is not already a foregone conclusion.

In conclusion, I would like to thank the participants in the Syracuse
Relativity Tea for stimulating comments on these topics, and R.B. Salgado
in particular for suggesting the most appropriate formulation of the
lemma utilized above.
This research was partly supported by NSF grant PHY 9307570.

\vskip 0.5truein
\centerline {\bf References}
\nobreak
\medskip
\noindent
\parindent=0pt
\parskip=3pt
\singlespace

[1]
 Sorkin, R.D.,
  ``Consequences of Spacetime Topology'',
     in {\it Proceedings of the Third Canadian Conference on General
          Relativity and Relativistic Astrophysics},
           (Victoria, Can\-ada, May 1989),
             edited  by A. Coley, F. Cooperstock and B. Tupper, 137-163
               (World Scientific, 1990)

[2]
 Bombelli, L., Lee, J., Meyer, D. and R.D. Sorkin,
``Spacetime as a Causal Set'',
  {\it Phys. Rev. Lett.} {\bf 59}, 521-524 (1987);
  Sorkin, R.D.,
    ``Spacetime and Causal Sets'',
          in J.C. D'Olivo, E. Nahmad-Achar, M. Rosenbaum, M.P. Ryan,
              L.F. Urrutia and F. Zertuche (eds.),
          {\it Relativity and Gravitation:  Classical and Quantum,}
          (Proceedings of the {\it SILARG VII Conference},
            held Cocoyoc, Mexico, December, 1990), pages 150-173,
             (World Scientific, Singapore, 1991)

[3]
Bialynicki-Birula, Iwo,
``Transition amplitudes versus transition probabilities and a
  reduplication of space-time'',
  in {\it Quantum Concepts in Space and Time},
     R. Penrose and C.J. Isham (eds.)
    (Oxford, Clarendon, 1986) pp. 226-235

[4]
 Sorkin, R.D.,
``On the Role of Time in the Sum-over-histories Framework for Gravity'',
  in the proceedings of the conference ``The History of Modern
     Gauge Theories'', held Logan, Utah, July, 1987;
       to be published in
         {\it Int. J. Theor. Phys.} (1994, to appear)

[5]
Sinha, Sukanya and R.D Sorkin
 ``A Sum-Over-Histories-Account of an EPR(B) Experiment''
    {\it Found. of Phys. Lett.}, {\bf 4}, 303-335, (1991)

[6]
 Sorkin, R.D.,
``Problems with Causality in the Sum-over-histories
           Framework for Quantum Mechanics'', in A. Ashtekar and J. Stachel
           (eds.), {\it Conceptual Problems of Quantum Gravity} (Proceedings
           of the conference of the same name, held Osgood Hill, Mass., May
           1988), 217--227 (Boston, Birkh\"auser, 1991)

[7]
R.D. Sorkin, ``Forks in the road, on the way to quantum
     gravity'', talk presented  at the Brill-Misner Fest, College Park,
     Maryland 1993, Syracuse University Preprint: SU-GP-93-12-2

[8]
R.D. Sorkin, in preparation.

[9]
Hartle, J.B., ``The Quantum Mechanics of Cosmology'', in {\it Quantum
 Cosmology and Baby Universes: Proceedings of the 1989 Jerusalem Winter
 School for Theoretical Physics}, eds. S. Coleman et al. (World
 Scientific, Singapore, 1991)

[10]
R. Geroch, ``The Everett interpretation", {\it No\^{u}s} {\bf 18}:617 (1984).

[11]
Isham, C.J.,
``Quantum Logic and the Histories Approach to Quantum Theory'',
  Imperial College Preprint: Imperial/TP/92-93/39,
  (gr-qc/9308006)

\end